\documentclass[12pt,fleqn,showpacs,preprintnumbers,amsmath,amssymb]{article}
\usepackage{graphicx}
\usepackage{amssymb}
\usepackage{amsmath}
\makeatletter \oddsidemargin -0.1in \evensidemargin -.1in
\newcommand{\singlespacing}{\let\CS=\@currsize\renewcommand{\baselinestretch}{2}\tiny\CS}
\newcommand{\doublespacing}{\let\CS=\@currsize\renewcommand{\baselinestretch}{2}\tiny\CS}
\begin{document}
\title{Dust ion acoustic solitary structures in presence of nonthermally distributed electrons and positrons}
\author{Ashesh Paul$^{1}$, Anup Bandyopadhyay$^{1*}$ and K. P. Das$^{2}$\\
{\small{\textit{\ $^{1}$Department of Mathematics, Jadavpur University, Kolkata - 700 032, India.}}}\\
{\ $^{*}$email - abandyopadhyay1965@gmail.com}\\
{\small{\textit{\ $^{2}$Department of Applied Mathematics, University of Calcutta,}}}\\
{\small{\textit{\ 92 Acharya Prafulla Chandra Road, Kolkata - 700009, India.}}}\\}

%
%

\date{}
\maketitle
\begin{abstract}
\noindent The purpose of this paper is to extend the recent work of Paul \& Bandyopadhyay [\textit{Astrophys. Space Sci.} \textbf{361}, 172(2016)] on the existence of different dust ion acoustic solitary structures in an unmagnetized collisionless dusty plasma consisting of negatively charged static dust grains, adiabatic warm ions, nonthermal electrons and isothermal positrons in a more generalized form by considering nonthermal positrons instead of isothermal positrons. The present system supports both positive and negative potential double layers, coexistence of solitary waves of both polarities and positive potential supersolitons. The qualitative and the quantitative changes in existence domains of different solitary structures which occur for the presence of nonthermal positrons have been presented in comparison with the results of Paul \& Bandyopadhyay [\textit{Astrophys. Space Sci.} \textbf{361}, 172(2016)]. The formation of supersoliton structures and their limitations have been analyzed with the help of phase portraits of the dynamical system corresponding to the dust ion acoustic solitary structures. Phase portrait analysis clearly indicates a smooth transition between soliton and supersoliton.
\end{abstract}

\noindent \textbf{PACS :} 52.27.Lw, 52.35.Fp, 52.35.Mw, 52.35.Sb

\noindent \textbf{Keywords :} Dust ion acoustic waves, Sagdeev pseudo-potential, Nonthermal electrons and positrons, Solitons, Double layers, Supersolitons, Phase portrait.
\section{Introduction}

Positrons are present in various dusty astrophysical environments such as in the remnants of supernova explosions which can last for thousand of years in space \cite{alfven1981}, around the pulsars \cite{shukla04}, near the surface of the neutron stars \cite{zeldovich1971,shukla04}, in the hot spots on ``dust ring'' in the galactic centre \cite{zurek1985}, interstellar medium \cite{zurek1985,higdon09,shukla2008}, interior regions of accretion disks near neutron stars and magnetars \cite{dubinov12}, in Milky way \cite{shukla2008} etc. Therefore, the plasma system of these astrophysical sites eventually make four component electron-positron-ion-dust (e-p-i-d) plasma. Also, e-p-i-d plasma may be present in the magnetosphere and in the ionosphere of the Earth as such regions of the atmosphere of the Earth contain highly charged dust grains \cite{alfven1981} and positrons \cite{gusev2000,gusev2001}. 
Positrons are also found in some well known astrophysical dusty plasma environments such as in the magnetosphere of the Jupiter \cite{merlino2006} and the Saturn \cite{horanyi2004}. The existence of e-p-i-d plasma in such numerous cosmic sites and also in laboratory environments \cite{shukla04,dubinov12} motivate the researchers to investigate nonlinear wave structures in such plasmas.

Several authors \cite{ghosh08,dubinov12,el-tantawy11a,el-tantawy11b,saini13} investigated small or arbitrary amplitude dust ion acoustic (DIA) solitary structures in different e-p-i-d plasma system consisting of isothermal positrons. In some cosmic sites, the velocity distribution functions of charged particles are not only non-Maxwellian but also highly anisotropic with an excess of high energy particles \cite{alfven1981}. Now, positron being the antimatter of electron, has the same mass and charge of an electron but of opposite sign. Therefore, positrons can be considered as a lighter species of a plasma and may acquire energy as high as electrons. Therefore, in cosmic plasma, the existence of fast energetic positrons is very natural. Thus the investigation of DIA solitary structures in presence of Cairns distributed \cite{cairns95} nonthermal positrons is instructive and also is an unexplored field till date.

Recently, Paul \& Bandyopadhyay \cite{paul2016} investigated the arbitrary amplitude DIA solitary structures in an e-p-i-d plasma consisting of nonthermal electrons and isothermal positrons and reported the existence of double layers of both polarities, coexistence of solitary waves of both polarities and most importantly the existence of positive potential supersolitons \cite{dubinov12b,verheest13a,verheest13b,verheest13c,verheest14,verheest14a}. The present paper is an extension of the recently published paper of Paul \& Bandyopadhyay \cite{paul2016} in the following directions.\\ 
(i) Instead of isothermal positrons, the Cairns distributed \cite{cairns95} nonthermal positrons has been considered.\\
(ii) The qualitative and the quantitative changes in the existence domains of different solitary structures which occur for the presence of nonthermal positrons have been discussed in comparison with the results of Paul \& Bandyopadhyay \cite{paul2016}.\\
(iii) The existence of different DIA solitary structures has been investigated when the nonthermal parameter associated with the nonthermal velocity distribution function of positrons is same as that of electrons.\\
(iv) Another important aspect of this paper is to analyze the formation of supersoliton structures and their limitations with the help of phase portraits of the dynamical system corresponding to the DIA solitary structures.\\
(v) For the first time, the transition from supersoliton to soliton has been explained through the phase portrait analysis of the dynamical system corresponding to the solitary structures. 

This paper is organized as follows: the basic equations and the energy integral are given in \S \ref{sec:basic_eqn}. In \S \ref{sec:solution_spaces}, DIA solitary structures have been thoroughly presented with the help of the qualitatively distinct compositional parameter spaces.  The phase portraits of the dynamical system corresponding to the DIA solitary structures have been analyzed in \S \ref{sec:Phase_Portraits}. Finally, conclusions are given in \S \ref{sec:Conclusions}.

\section{\label{sec:basic_eqn}Basic Equations \& Energy Integral}
We consider a collisionless unmagnetized multicomponent dusty plasma system consisting of adiabatic warm ions, negatively charged static dust particulates, nonthermally distributed electrons and positrons. The nonlinear behaviour of DIA waves propagating along $x$-axis may be described by the following set of equations consisting of the ion continuity equation, ion fluid equation of motion, the pressure equation for ion fluid and the Poisson equation.
\begin{eqnarray}\label{continuity}
\frac{\partial n_{i}}{\partial t}+\frac{\partial}{\partial x}(n_{i}u_{i})=0,
\end{eqnarray}
\begin{eqnarray}\label{momentum}
M_{s}^{2}\bigg(\frac{\partial u_{i}}{\partial t}+u_{i}\frac{\partial u_{i}}{\partial x}\bigg)+\frac{(1-p)\sigma_{ie}}{n_{i}}\frac{\partial p_{i}}{\partial x}+\frac{\partial \phi}{\partial x}=0,
\end{eqnarray}
\begin{eqnarray}\label{pressure}
\frac{\partial p_{i}}{\partial t}+u_{i}\frac{\partial p_{i}}{\partial x}+\gamma p_{i} \frac{\partial u_{i}}{\partial x}=0,
\end{eqnarray}
\begin{eqnarray}\label{poisson}
\frac{\partial^{2} \phi}{\partial x^{2}}=-\frac{M_{s}^{2}-\gamma \sigma_{ie}}{1-p}\bigg(n_{i}-n_{e}+n_{p}-\frac{Z_{d}n_{d0}}{n_{0}}\bigg).
\end{eqnarray}
Here $n_{i}$, $n_{e}$, $n_{p}$, $u_{i}$, $p_{i}$, $\phi$, $x $ and $ t $ are, respectively, number density of ion, number density of electron, number density of positron, velocity of ion fluid, ion fluid pressure, electrostatic potential, spatial variable and time, and these have been normalized by $n_{0}$ ($=n_{i0}+n_{p0}=n_{e0}+Z_{d}n_{d0}$), $n_{0}$, $n_{0}$, $C_{D}$ (linearized velocity of the DIA wave in the present plasma system for long-wavelength plane wave perturbation), $n_{i0}K_{B}T_{i}$, $\Phi=\frac{K_{B}T_{e}}{e}$, $ \lambda_{D} $ (Debye length of the present plasma system) and $\lambda_{D}/C_{D}$ with $n_{e0}$, $n_{i0}$, $n_{p0}$ and $n_{d0}$ are, respectively, the equilibrium number densities of electrons, ions, positrons and dust particulates, $ \gamma(=3) $ is the adiabatic index, $ Z_{d} $ is the number of electrons residing on a dust grain surface, $-e$ is the charge of an electron, $T_{i}$ ($T_{e}$) is the average temperature of ions (electrons) and $K_{B}$ is the Boltzmann constant. The expressions of $M_{s}$ and the four basic parameters $p$, $\mu$, $\sigma_{ie}$, $\sigma_{pe}$ are given by the following equations: 
\begin{eqnarray}\label{Ms}
M_{s}=\sqrt{\gamma\sigma_{ie}+\frac{(1-p)\sigma_{pe}}{p(1-\beta_{p})+\mu (1-\beta_{e}) \sigma_{pe}}},	
\end{eqnarray}
\begin{eqnarray}\label{p}
p=\frac{n_{p0}}{n_{0}},~\mu=\frac{n_{e0}}{n_{0}},~\sigma_{ie}=\frac{T_{i}}{T_{e}},~\sigma_{pe}=\frac{T_{p}}{T_{e}},	
\end{eqnarray}
where $T_{p}$ is the average temperature of positrons, $\beta_{e}$ and $\beta_{p}$ are, respectively, the nonthermal parameters associated with the Cairns model for electron and positron species, and according to  Verheest \& Pillay \cite{verheest08}, the physically admissible bounds of $\beta_{e}$ and $\beta_{p}$ are given by $0 \leq \beta_{e},~ \beta_{p} \leq \frac{4}{7} \approx 0.6$.   

It is important to note that the equations (\ref{momentum}) and (\ref{poisson}) of the present paper are not same as the equatios (2) and (4) of Paul \& Bandyopadhyay \cite{paul2016} because the expression of $M_{s}^{2}$ as given in the equation (\ref{Ms}) is not same as the equation (5) of Paul \& Bandyopadhyay \cite{paul2016}. Although, the equations (\ref{continuity}) - (\ref{poisson}) of the present paper are, respectively, same as the equations (1) - (4) of Paul \& Bandyopadhyay \cite{paul2016} if we put $\beta_{p}=0$.
   
Under the above-mentioned normalization of the dependent variables, the number density of nonthermally distributed electrons and positrons are, respectively, given by
\begin{eqnarray}\label{ne}
n_{e} = \mu(1-\beta_{e}\phi+\beta_{e}\phi^{2})e^{\phi},
\end{eqnarray}
\begin{eqnarray}\label{np}
n_{p} = p\bigg[1+\beta_{p}\bigg(\frac{\phi}{\sigma_{pe}}\bigg)+\beta_{p}\bigg(\frac{\phi}{\sigma_{pe}}\bigg)^{2}\bigg] e^{-\phi / \sigma_{pe}},
\end{eqnarray}
The above equations are supplemented by the following unperturbed charge neutrality condition
\begin{eqnarray}
n_{i0}+n_{p0}=n_{e0}+Z_{d}n_{d0}.
\end{eqnarray} 

To  investigate the steady state arbitrary amplitude DIA solitary structures, we consider the transformation $ \xi=x-Mt $, where $M$ is the dimensionless velocity of the wave frame normalized by the linearized DIA speed ($C_{D}$) for long-wavelength plane wave perturbation. Using this transformation and applying the boundary conditions:\\ $ \big(n_{i},p_{i},u_{i},\phi,\frac{d\phi}{d\xi}\big)\rightarrow \big(1-p,1,0,0,0\big)\mbox{    as    }  |\xi|\rightarrow \infty,
$\\ we get the following energy integral:
\begin{eqnarray}\label{energy_integral}
\frac{1}{2}\bigg(\frac{d\phi}{d\xi}\bigg)^{2}+V(\phi)=0,
\end{eqnarray}
where
\begin{eqnarray}\label{V_phi_1}
V(\phi) = (M_{s}^{2}-3\sigma_{ie}) \Big[~V_{i} +\frac{p }{1-p} \sigma_{pe} V_{p}-\frac{\mu}{1-p}V_{e} -\frac{1-\mu}{1-p}V_{d}\Big],
\end{eqnarray}
\begin{eqnarray}\label{V_i_1}
V_{i} = M^{2}M_{s}^{2}+\sigma_{ie} -N_{i}\Big[M^{2}M_{s}^{2}+3\sigma_{ie}-2\phi  -2\sigma_{ie}N_{i}^{2}\Big],
\end{eqnarray}
\begin{eqnarray}\label{N_i_1}
N_{i}=\frac{n_{i}}{1-p}=\frac{MM_{s}\sqrt{2}}{(\sqrt{\Phi_{M}-\phi}+\sqrt{\Psi_{M}-\phi})},
\end{eqnarray}
\begin{eqnarray}\label{Phi_M_1}
\Phi_{M} = \frac{1}{2}\Big(MM_{s}+\sqrt{3\sigma_{ie}}\Big)^{2}, 
\end{eqnarray}
\begin{eqnarray}\label{Psi_M_1}
\Psi_{M} = \frac{1}{2}\Big(MM_{s}-\sqrt{3\sigma_{ie}}\Big)^{2},
\end{eqnarray}
\begin{eqnarray}\label{V_e_1}
V_{e} = \big(1+3\beta_{e}-3\beta_{e}\phi+\beta_{e}\phi^{2}\big)e^{\phi}-(1+3\beta_{e}),
\end{eqnarray}
\begin{eqnarray}\label{V_p_1}
V_{p} = (1+3\beta_{p}) -\bigg[1+3\beta_{p} + 3\beta_{p}\bigg(\frac{\phi}{\sigma_{pe}}\bigg) 
+\beta_{p}\bigg(\frac{\phi}{\sigma_{pe}}\bigg)^{2}\bigg]e^{-\phi / \sigma_{pe}},
\end{eqnarray}
\begin{eqnarray}\label{V_d_1}
V_{d}=\phi.
\end{eqnarray}
According to Sagdeev \cite{sagdeev66}, for the existence of a positive (negative) potential solitary wave [PPSW] ([NPSW]) solution of (\ref{energy_integral}), we must have the following three conditions: 
(i) $\phi=0$ is the position of unstable equilibrium of a particle of unit mass associated with the energy integral (\ref{energy_integral}), i.e., $V(0)=V'(0)=0$ and $V''(0)<0$.
(ii) $V(\phi_{m}) = 0$, $V'(\phi_{m}) > 0$ ($V'(\phi_{m}) < 0$) for some $\phi_{m} > 0$ ($\phi_{m} < 0$). This condition is responsible for the oscillation of the particle within the interval $\min\{0,\phi_{m}\}<\phi<\max\{0,\phi_{m}\}$.
(iii) $V(\phi) < 0$ for all $0 <\phi < \phi_{m}$ ($\phi_{m} < \phi < 0$). This condition is necessary to define the energy integral (\ref{energy_integral}) within the interval $\min\{0,\phi_{m}\}<\phi<\max\{0,\phi_{m}\}$. For the existence of a positive (negative) potential double layer [PPDL] ([NPDL]) solution of (\ref{energy_integral}), the second condition is replaced by $V(\phi_{m}) = 0$, $V'(\phi_{m}) = 0$, $V''(\phi_{m}) < 0$ for some $\phi_{m} > 0$ ($\phi_{m} < 0)$. This condition states that the particle cannot be reflected back from the point $\phi=\phi_{m}$ to the point $\phi = 0$.

Using the condition (i) we get, $M>M_{c}=1$, i.e., the solitary structures start to exist just above the curve $M = M_{c}=1$. From the equation (\ref{N_i_1}), we see that $ N_{i} $ is real if and only if $ \phi \leq \Psi_{M}$. Using this condition and following the similar argument as given in the papers of Das \textit{et al.} \cite{das09,das12}, we get an upper bound $M_{max}$ of  the  Mach  number $ M $ for the existence of all PPSWs. Here, $ M_{max} $ is the largest positive root of the equation $ V(\Psi_{M}) = 0 $ subject to the condition $ V(\Psi_{M}) \geq 0 $ for all $ M \leq M_{max} $. Therefore, $M$ assumes its upper limit $ M_{max} $ for the existence of all PPSWs when $\phi$ tends to $\Psi_{M}$, i.e., when ion number density goes to maximum compression. 

Using the conditions for the existence of NPDL (PPDL) and following the same argument of Das \textit{et al.} \cite{das12}, we can easily develop an algorithm to find the Mach number $M_{NPDL}$ ($M_{PPDL}$) corresponding to a NPDL (PPDL) solution of the energy integral (\ref{energy_integral}). Now if both $M_{max}$ and $M_{PPDL}$ exist, then we have $M_{c}<M_{PPDL} < M_{max}$ and for $M_{PPDL}<M \leq M_{max}$, we get PPSWs after the formation of PPDL, and consequently, the existence of positive potential supersolitons (PPSSs) (according to  Dubinov \& Kolotkov \cite{dubinov12b}) is confirmed.

\section{\label{sec:solution_spaces} Different existence domains}
\begin{figure}
\begin{center}
  \includegraphics{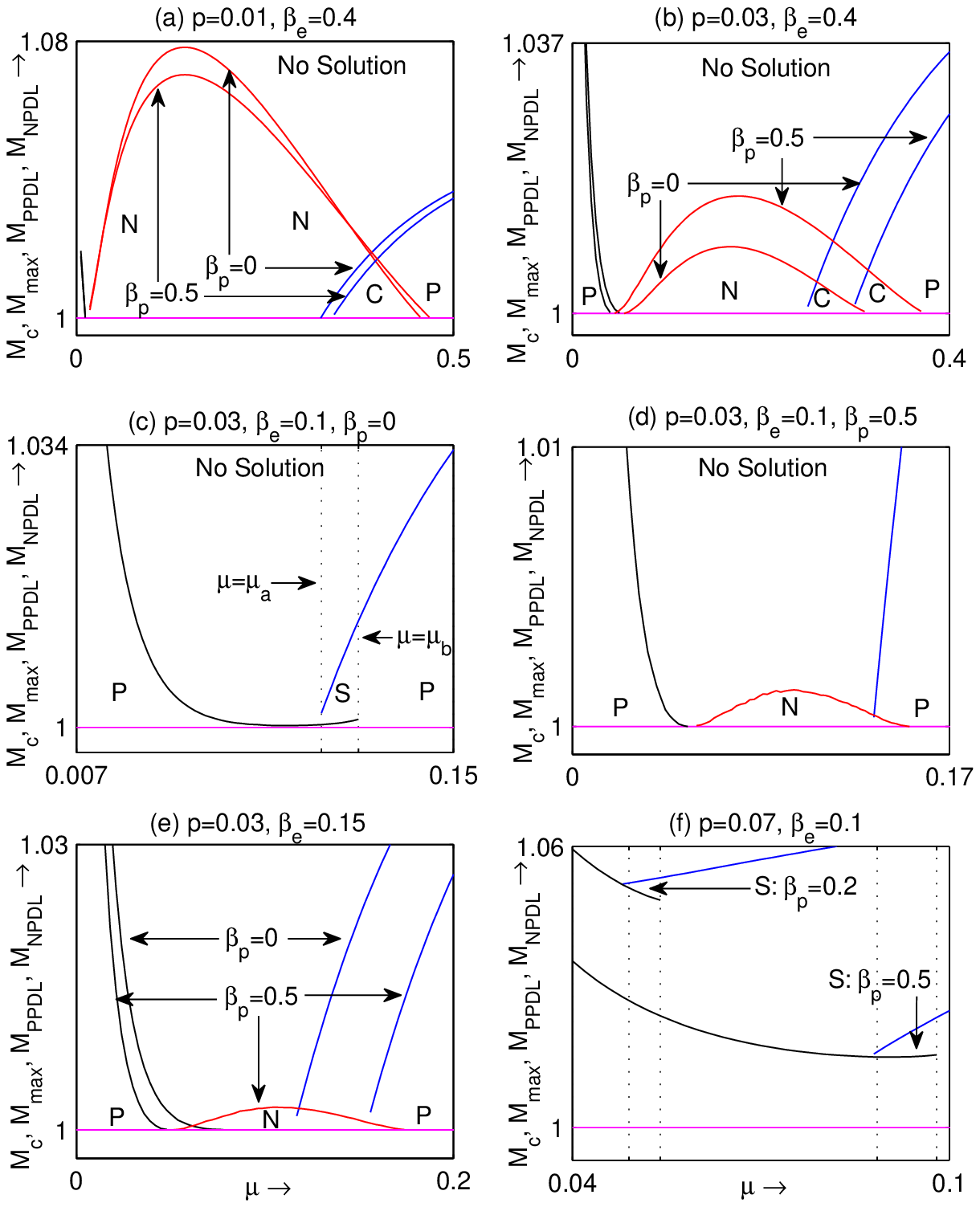}
  \caption{\label{fig_sol_spc_wrt_mu} Existence domains with respect to $\mu$ for different values of parameters as shown in the figures with $\sigma_{ie}=\sigma_{pe}=0.9$. The red curve, the magenta curve, the blue curve and the black curve correspond to the curves $M=M_{NPDL}$, $M=M_{c}$, $M=M_{max}$ and $M=M_{PPDL}$ respectively.}
  \end{center}
\end{figure}

\begin{figure}
\begin{center}
  \includegraphics{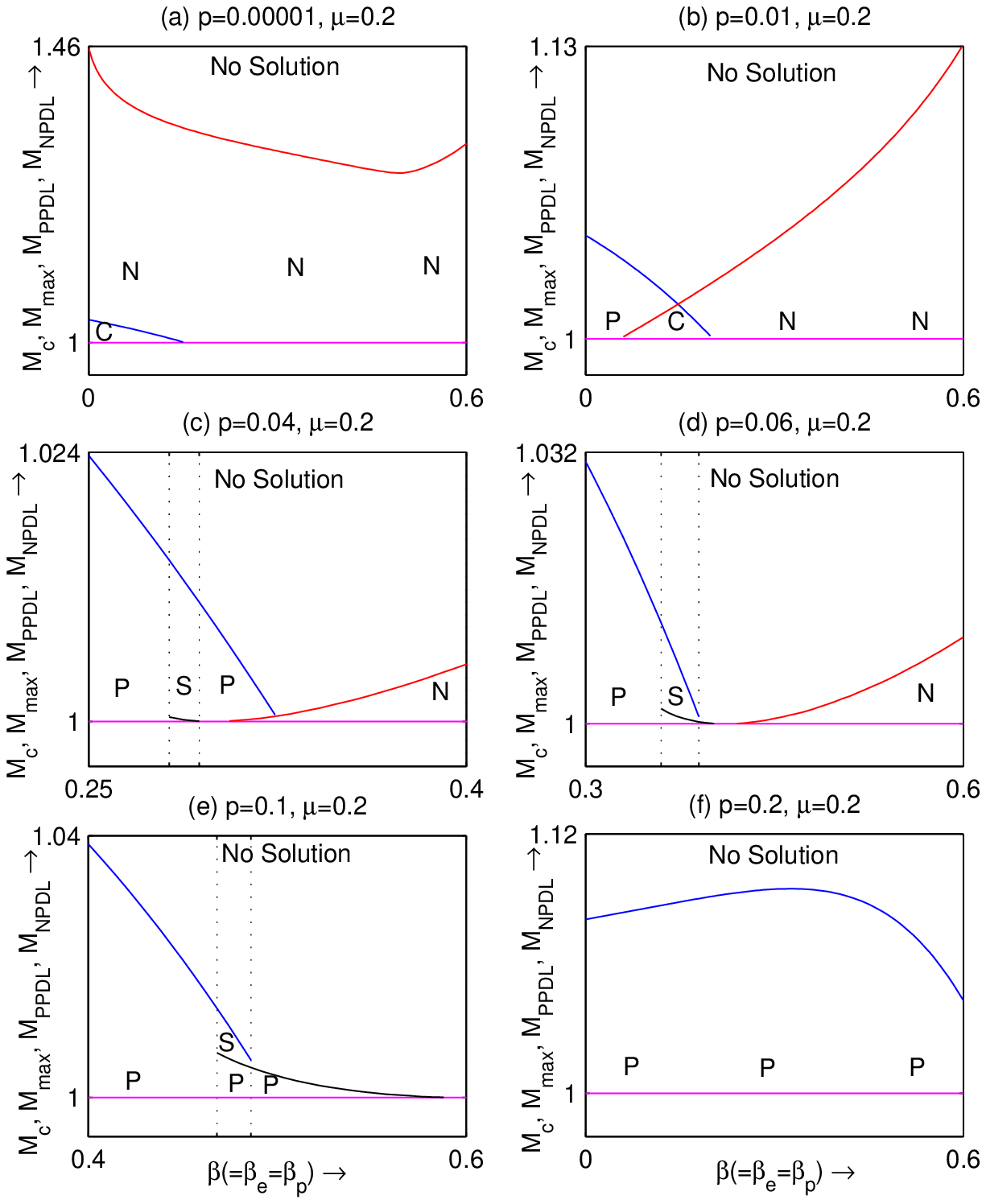}
  \caption{\label{fig_sol_spc_wrt_beta} Existence domains with respect to $\beta(=\beta_{e}=\beta_{p})$ for different values of parameters as shown in the figures with $\sigma_{ie}=\sigma_{pe}=0.9$. The red curve, the magenta curve, the blue curve and the black curve correspond to the curves $M=M_{NPDL}$, $M=M_{c}$, $M=M_{max}$ and $M=M_{PPDL}$ respectively.}
\end{center}
\end{figure}

Figure \ref{fig_sol_spc_wrt_mu}(a) - figure \ref{fig_sol_spc_wrt_mu}(f) are the qualitatively different existence domains with respect to $\mu$ for different values of positron concentration $p$ and the nonthermal parameters associated with the distribution functions of electrons and positrons. Figure \ref{fig_sol_spc_wrt_beta}(a) - figure \ref{fig_sol_spc_wrt_beta}(f) are the existence domains with respect to $\beta(=\beta_{e}=\beta_{p})$ for different values of $p$. In the above mentioned figures, P, N, S and C denote, respectively, the existence regions of PPSWs, NPSWs, PPSWs after the formation of PPDL and the region of coexistence of both PPSWs and NPSWs.
 
These figures are self-explanatory. For example, from figure \ref{fig_sol_spc_wrt_beta}(c), we see that (i) the system supports PPSWs which are restricted by $M_{c}<M \leq M_{max}$. (ii) The system supports NPDLs along the curve $M=M_{NPDL}$ and consequently, it supports NPSWs and the existence region of NPSWs is bounded by the curves $M=M_{c}$ and $M=M_{NPDL}$. (iii) In between the cut-off values $\beta^{(a)}(=0.282)$ and $\beta^{(b)}(=0.294)$ of $\beta$, the system supports PPDL along the curve $M=M_{PPDL}$. But in this interval of $\beta$ ($\beta^{(a)}<\beta<\beta^{(b)}$), we have $M_{PPDL}<M_{max}$, i.e., there exist PPSWs after the formation of PPDLs if the Mach number $M$ is restricted by $M_{PPDL}<M \leq M_{max}$. Consequently, in this region, the existence of PPSSs is confirmed.

In figure \ref{fig_sol_spc_wrt_mu}(a) we see that there is no qualitative change in the existence domain with respect to $\mu$ for the variation of $\beta_{p}$. The existence region of NPSWs bounded by the curves $M=M_{c}$ and $M=M_{NPDL}$ and the existence region of PPSWs bounded by the curves $M=M_{c}$ and $M=M_{max}$,  both decrease with increasing $\beta_{p}$. On the other hand, figure \ref{fig_sol_spc_wrt_mu}(b) shows that for increasing $\beta_{p}$, the existence region of PPSWs bounded by the curves $M=M_{c}$ and $M=M_{max}$ decreases whereas the existence region of NPSWs increases.

From figure \ref{fig_sol_spc_wrt_mu}(c) we see that whenever $\beta_{p}=0$, the system supports PPSSs whereas it does not support any negative potential solitary structure (see also figure 4(a) of Paul \& Bandyopadhyay \cite{paul2016}). But with the increment in $\beta_{p}$, the existence region of PPSSs decreases and if $\beta_{p}$ exceeds a cut-off value then the system does not support any PPSS. For further increment in $\beta_{p}$, we see that the system starts to support NPDLs along the curve $M=M_{NPDL}$ if $\beta_{p}$ exceeds another cut-off value. Consequently, it supports NPSWs. There also exists a region of coexistence of solitary waves of both polarities and this coexistence region is bounded by the curves $M=M_{c}$, $M=M_{NPDL}$ and $M=M_{max}$ (see figure \ref{fig_sol_spc_wrt_mu}(d)).

Now, for isothermal positrons, the system does not support any negative potential solitary structure for any $\mu$ with $\beta_{e}=0.15$, $p=0.03$ and $\sigma_{ie}=\sigma_{pe}= 0.9$ (see figure 6(a) of Paul \& Bandyopadhyay \cite{paul2016}). But for nonthermal positrons, the system supports NPDLs along the curve $M=M_{NPDL}$ and consequently, it supports NPSWs and coexistence of solitary waves of both polarities (see figure \ref{fig_sol_spc_wrt_mu}(e)). For increasing $\beta_{p}$, the region of existence of NPSWs increases  whereas the existence region of PPSWs decreases.

Figure 8(a) of Paul \& Bandyopadhyay \cite{paul2016} shows that for isothermal positrons, the system does not support any PPSS for any physically admissible value of $\mu$ when $\beta_{e}=0.1$, $p=0.07$ and $\sigma_{ie}=\sigma_{pe}= 0.9$. But figure \ref{fig_sol_spc_wrt_mu}(f) of the present paper shows that for nonthermal positrons the system supports PPSSs and the region of existence of PPSSs increases with increasing $\beta_{p}$. Similarly, for isothermal positrons, the system does not support PPSSs (see figure 10 of Paul \& Bandyopadhyay \cite{paul2016}) whereas for nonthermal positrons the system supports PPSSs and the region of existence of PPSSs increases with increasing $\beta_{p}$ for $\beta_{e}=0.3$, $p=0.2$, $\sigma_{ie}=\sigma_{pe}= 0.9$ and for any $\mu>0$. 

Now, we discuss the existence domains with respect to $\beta(=\beta_{e}=\beta_{p})$ for a fixed value of $\mu$ and different values of $p$ starting from $p=0.00001$. Although the figure \ref{fig_sol_spc_wrt_beta}(a) corresponds to the existence domain for $p=0.00001$ with $\mu=0.2$ but qualitatively it represents the existence domain or compositional parameter space for any $p$ lying within $0 < p \leq p^{(a)}$. Similarly, figure \ref{fig_sol_spc_wrt_beta}(b), \ref{fig_sol_spc_wrt_beta}(c), \ref{fig_sol_spc_wrt_beta}(d) and figure \ref{fig_sol_spc_wrt_beta}(e) represent the existence domains for any $p$ lying within $p^{(a)} < p \leq p^{(b)}$, $p^{(b)} < p \leq p^{(c)}$, $p^{(c)} < p \leq p^{(d)}$ and $p^{(d)} < p \leq p^{(e)}$ respectively. Finally, figure \ref{fig_sol_spc_wrt_beta}(f) stands for the existence domain for any $p>p^{(e)}$. For $\mu=0.2$ with $\sigma_{ie}=\sigma_{pe}= 0.9$, the values of $p^{(a)}$, $p^{(b)}$, $p^{(c)}$, $p^{(d)}$ and $p^{(e)}$ are, respectively, 0.003, 0.026, 0.047, 0.097 and 0.182.

Thus figures \ref{fig_sol_spc_wrt_beta}(a) - \ref{fig_sol_spc_wrt_beta}(f) show that for a fixed value of $\mu$ whenever $p$ lies in the interval $0<p \leq p^{(a)}$, the system supports NPDLs along the curve $M=M_{NPDL}$ for any physically admissible value of $\beta(=\beta_{e}=\beta_{p})$. The system also supports coexistence of solitary waves of both polarities for a certain interval of $\beta$ as shown in figure \ref{fig_sol_spc_wrt_beta}(a). Now, in this interval of $p$, if we increase $p$, the existence region of NPSWs decreases whereas the existence region of PPSWs bounded by the curves $M=M_{c}$ and $M=M_{max}$ increases. Thus, for $p^{(a)} < p \leq p^{(b)}$, we have two cut-off values of $\beta$, viz., $\beta_{1}$ and $\beta_{2}$ such that the system supports only PPSWs in $0<\beta \leq\beta_{1}$, coexistence of both NPSWs and PPSWs in $\beta_{1}<\beta \leq \beta_{2}$ and only NPSWs for $\beta > \beta_{2}$ (see figure \ref{fig_sol_spc_wrt_beta}(b)). For $p^{(b)} < p \leq p^{(c)}$, the system starts to support PPDL along the curve $M=M_{PPDL}$ whenever $\beta$ lies in the interval $\beta^{(a)}(=0.282)<\beta<\beta^{(b)}(=0.294)$. Furthermore, in $\beta^{(a)}<\beta<\beta^{(b)}$, $M_{PPDL}<M_{max}$ and consequently, in this region of parameter space, the system supports PPSWs after the formation of PPDL and hence the existence of PPSSs is confirmed (see figure \ref{fig_sol_spc_wrt_beta}(c)). For $p^{(c)}<p\leq p^{(d)}$, although the system supports PPSWs, NPSWs, PPDLs, NPDLs and PPSSs but it does not support coexistence of solitary waves of both polarities (see figure \ref{fig_sol_spc_wrt_beta}(d)). In this interval of $p$, i.e., in $p^{(c)}<p\leq p^{(d)}$, if we again increase $p$, then the existence region of NPSWs decreases whereas the existence region of PPSWs bounded by the curves $M=M_{c}$ and $M=M_{PPDL}$ increases. For $p^{(d)} < p \leq p^{(e)}$, the system does not support negative potential solitary structures anymore but it supports PPDL and $M_{PPDL}<M_{max}$ (see figure \ref{fig_sol_spc_wrt_beta}(e)). Again, in this interval of $p$, the existence region of PPSWs bounded by the curves $M=M_{c}$ and $M=M_{PPDL}$ decreases for increasing $p$ and as a result, the existence region of the positive potential solitons after the formation of double layer decreases and finally, disappears from the system for $p> p^{(e)}$. Figure \ref{fig_sol_spc_wrt_beta}(f) shows that for $p>p^{(e)}$, the system supports only PPSWs bounded by the curves $M=M_{c}$ and $M=M_{max}$ for any physically admissible value of $\beta$.


\section{\label{sec:Phase_Portraits} Phase Portraits of different solitary structures}

Differentiating the energy integral (\ref{energy_integral}) with respect to $\phi$, we get
\begin{eqnarray}\label{energy_integral_differentiation}
\frac{d^{2}\phi}{d\xi^{2}}+V'(\phi)=0.
\end{eqnarray}
This equation is equivalent to the following system of differential equations
\begin{eqnarray}\label{phase_portraits}
\frac{d\phi_{1}}{d\xi}=\phi_{2}~,~\frac{d\phi_{2}}{d\xi}=-V'(\phi_{1})~,
\end{eqnarray}
where $\phi_{1}=\phi$. In the present paper, we have considered the supersoliton structures that occur beyond double layers with the help of qualitatively different existence domains. Now, we explain their different unusual shapes with the help of phase portraits of the system of coupled equations (\ref{phase_portraits}) in the $\phi_{1}-\phi_{2}$ plane.

To describe the existence and the unusual shape of PPSSs, we consider figure \ref{pp1_soliton_epsilon=0_pt_001} - figure \ref{pp1_supersoliton_epsilon=0_pt_0017} and figure \ref{saddle_points}, where we have used the existence domain as shown in figure \ref{fig_sol_spc_wrt_mu}(f) to determine the Mach numbers for the formation of PPSSs. To draw the phase portraits of the coexistence of solitons of both polarities as shown in  figure \ref{pp1_coexistence}, we have used the existence domain with respect to $\mu$ for $p=0.01$, $\beta_{e}=0.1$ and $\beta_{p}=0.1$ with $\sigma_{ie}=\sigma_{pe}=0.9$ to determine the Mach numbers for the coexistence of solitons of both polarities. 

In figure \ref{pp1_soliton_epsilon=0_pt_001} - figure \ref{pp1_supersoliton_epsilon=0_pt_0017} and in figure \ref{pp1_coexistence}, $V(\phi)$ is plotted against $\phi$ in the upper panel (or marked as (a)) of each figure. The lower panel (or marked as (b)) of each figure shows the phase portrait of the system (\ref{phase_portraits}). In these figures, we have used the values of the parameters as indicated in the figures with $\sigma_{pe}=\sigma_{ie}=0.9$. The curve $V(\phi)$ and the  phase portrait have been drawn on the same horizontal axis $\phi(=\phi_{1})$. The small solid circle corresponds to a saddle point whereas the small solid star indicates an equilibrium point other than saddle point of the system (\ref{phase_portraits}). It is simple to check that each maximum (minimum) point of $V(\phi)$ corresponds to a saddle point (an equilibrium point other than a saddle point) of the system (\ref{phase_portraits}).

From these figures, we see that there is a one-one correspondence between the separatrix of the phase portrait as shown with a heavy blue line in the lower panel with the curve $V(\phi)$ against $\phi$ of the upper panel. Again, it is important to note that the origin $(0,0)$ is always a saddle point of the system (\ref{phase_portraits}) and the separatrix corresponding to a solitary structure appears to start and end at the saddle point (0,0). The separatrix corresponding to a solitary structure is shown with a heavy blue line whereas other separatrices (if exist) are shown by green lines. The closed curve about an equilibrium point (other than a saddle point) contained in at least one separatrix indicates the possibility of the periodic wave solution about that fixed point. For example, the closed curves (red lines) of figure \ref{pp1_soliton_epsilon=0_pt_001}(b) about the fixed point (0.54,0) lying within the separatrix indicate the possibility of the periodic wave solutions about the fixed point (0.54,0).

Figure \ref{pp1_soliton_epsilon=0_pt_001}(a) shows the existence of a PPSW before the formation of PPDL and figure \ref{pp1_soliton_epsilon=0_pt_001}(b) shows that the corresponding phase portrait contains only one saddle at the origin and a non-zero equilibrium point. Consequently, there exists only one separatrix that appears to start and end at the origin enclosing a non-saddle fixed point. More precisely, the trajectory corresponding to the separatrix approaches the origin as $\xi \rightarrow \pm \infty$. It is also important to note that a separatrix corresponding to a solitary structure does not correspond to a periodic solution because for this case, the trajectory takes forever trying to reach a saddle point. In fact, this is the reason that a pseudo-particle associated with the energy integral (\ref{energy_integral}) takes an infinite long time to move away from its unstable position of equilibrium and then it continues its motion until $\phi$ takes the value $\phi_{m} (>0)$, where $V(\phi_{m})=0$ and $V'(\phi_{m})>0$ and again it takes an infinite long time to come back its unstable position of equilibrium \cite{verheest00}. Similarly, figure \ref{pp1_supersoliton_epsilon=0_pt_004} confirms the existence of a PPSW after the formation of PPDL.

From the phase portraits as given in figure \ref{pp1_soliton_epsilon=0_pt_001}(b) and figure \ref{pp1_supersoliton_epsilon=0_pt_004}(b), we see that there is no qualitative difference between these two phase portraits. Again, according to  Dubinov \& Kolotkov \cite{dubinov12b}, the separatrix corresponding to a supersoliton envelopes one or several inner separatrices and several equilibrium points. So, according to  Dubinov \& Kolotkov \cite{dubinov12b}, figure \ref{pp1_supersoliton_epsilon=0_pt_004}(b) does not correspond to a supersoliton. But figure \ref{phi_vs_xi_supersoliton} shows that there is a finite jump between the amplitudes of solitons before and after the formation of double layer. To explain this fact, we first of all consider the phase portrait corresponding to a double layer solution as given in figure \ref{pp1_ppdl}(b). Although figure \ref{pp1_ppdl}(a) shows that the curve $V(\phi)$ crosses the $\phi$-axis at the point $\phi=1.27$ (approximately) but if we enlarge the figure \ref{pp1_ppdl}(a) in the neighbourhood of the point $\phi=1.27$ then we see that $V(1.27)=0$, $V'(1.27)=0$ and $V''(1.27)<0$. Consequently, we have a double layer solution of the energy integral (\ref{energy_integral}) with $(0,~0)$ and $(1.27,~0)$ are two saddle points of the dynamical system (\ref{phase_portraits}). Figure \ref{pp1_ppdl}(b) shows that the separatrix corresponding to the double layer solution appears to pass through two saddle points and it encloses another two equilibrium points. If both the saddle points exist after a small increment of $M$ from $M=M_{PPDL}$ then the separatrix appears to pass through the origin encloses an inner sparatrix through a non-zero saddle and at least two equilibrium points as shown in the lower panel of figure \ref{pp1_supersoliton_epsilon=0_pt_0017}. Therefore, according to the definition of supersoliton (Dubinov and Kolotkov \cite{dubinov12b}), we see that for the same set of values of the parameters, $M=M_{PPDL}+0.0017$ defines a supersoliton whereas $M=M_{PPDL}+0.004$ does not define a supersoliton. But in both the cases we have a finite jump between the amplitudes of solitons after and before the formation of double layer. To make a clear difference between the solitary structures given in figure \ref{pp1_supersoliton_epsilon=0_pt_004}(b) and the lower panel of figure \ref{pp1_supersoliton_epsilon=0_pt_0017} for $M=M_{PPDL}+0.004$ and $M=M_{PPDL}+0.0017$ respectively, we consider figure \ref{saddle_points}. In this figure, we draw the saddle and other equilibrium points of the system (\ref{phase_portraits}) on the $\phi(=\phi_{1})$-axis for increasing values of $M$ starting from $M=M_{PPDL}+0.0001$. This figure shows that for increasing values of $M$ the distance between the non-zero saddle and the non-saddle fixed point nearest to the origin decreases and ultimately both of them disappear from the system. Finally, the system contains only one saddle at the origin and a non-zero equilibrium point. Consequently, only one separatrix enclosing the non-saddle fixed point is possible that appears to pass through the saddle at the origin. So, the existence of a soliton after the formation of a double layer confirms the existence of a sequence of supersolitons and there exists a critical value $M_{cr}$ of the Mach number $M$ such that for the existence of supersolitons after the formation of a double layer we must have $M_{PPDL}<M<M_{cr}$ whereas for $M_{cr} \leq M<M_{max}$, we get soliton like structures after the formation of a double layer. Thus, figure \ref{saddle_points} clearly shows the transition between soliton and supersoliton structures after the formation of a double layer. But it is important to note that there is always a finite jump between the amplitudes of solitons before and after the formation of double layer.    

From the lower panel of figure \ref{pp1_coexistence}, we see that there are two separatrices. The separatrix as shown by a heavy blue line corresponds the coexistence of solitons of both polarities and this separatrix is contained in another separatrix as shown with a green line. There exist infinitely many closed curves between these two separatrices and each of these closed curves corresponds to a super-nonlinear periodic wave as shown in figure 5(c) and figure 6(c) in the paper of  Dubinov \textit{et al.} \cite{dubinov12} for a dusty plasma system. However, further investigation of the super-nonlinear periodic wave solutions of the energy integral (\ref{energy_integral}) is beyond the scope of this paper.


\section{\label{sec:Conclusions} Conclusions}

In the present work, we have carried out a systematic investigation on the nature of existence of different DIA solitary structures in an unmagnetized dusty plasma consisting of negatively charged static dust grains, adiabatic warm ions, Cairns distributed nonthermal electrons and positrons with the help of existence domains and phase portraits. It is observed that the system supports PPDLs, NPDLs, coexistence of solitary waves of both polarities and PPSSs. All these solitary structures have also been observed when the nonthermal parameter associated with the velocity distribution function of positrons is same as that of electrons. The qualitative and the quantitative changes in the existence domains of different solitary structures which occur for the presence of nonthermal positrons have been discussed in comparison with the results of Paul \& Bandyopadhyay \cite{paul2016}. 

For fixed values of $p$, $\mu$, $\beta_{e}$, $\sigma_{ie}$ and $\sigma_{pe}$ we have the following observations. For increasing $\beta_{p}$, (i) the amplitude of NPSWs decreases (see figure \ref{fig_phi_vphi_new_new}(a)); (ii) the amplitude of NPDLs remains unchanged (see figure \ref{fig_phi_vphi_new_new}(b)); (iii) the amplitude of PPSWs bounded by the curves $M=M_{c}$ and $M=M_{max}$ increases (see figure \ref{fig_phi_vphi_new_new}(c)); (iv) the amplitude of PPDLs decreases (see figure \ref{fig_phi_vphi_new_new}(d)). (v) It can also be checked that the amplitude of PPSWs bounded by the curves $M=M_{c}$ and $M=M_{PPDL}$ decreases with increasing $\beta_{p}$.

\begin{figure}
\begin{center}
  \includegraphics{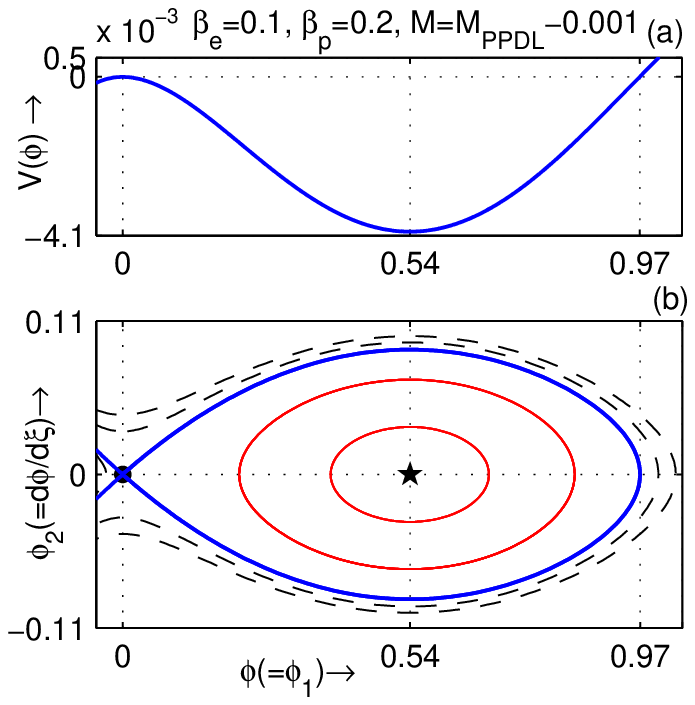}
  \caption{\label{pp1_soliton_epsilon=0_pt_001} $V(\phi)$ (on top) and the phase portrait of the system (\ref{phase_portraits}) (on bottom) have been drawn on the same $\phi(=\phi_{1})$-axis for $M=M_{PPDL}-0.001$ when $p=0.07$, $\mu=0.054$, $\beta_{e}=0.1$, $\beta_{p}=0.2$ and $\sigma_{ie}=\sigma_{pe}=0.9$.}
\end{center}
\end{figure}

\begin{figure}
\begin{center}
  \includegraphics{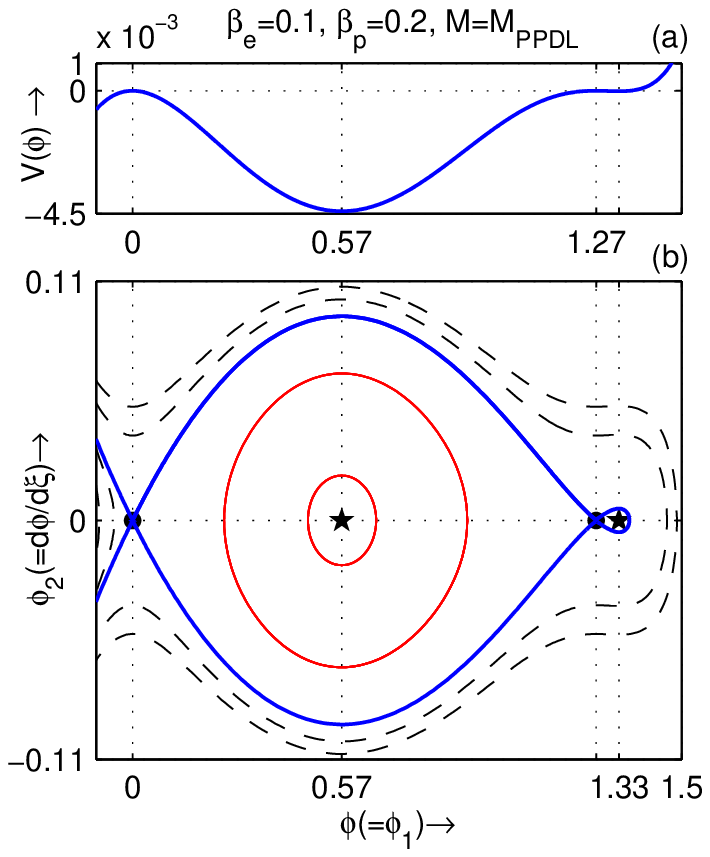}
  \caption{\label{pp1_ppdl} $V(\phi)$ (on top) and the phase portrait of the system (\ref{phase_portraits}) (on bottom) have been drawn on the same $\phi(=\phi_{1})$-axis for $M=M_{PPDL}$ when $p=0.07$, $\mu=0.054$, $\beta_{e}=0.1$, $\beta_{p}=0.2$ and $\sigma_{ie}=\sigma_{pe}=0.9$.}
\end{center}
\end{figure}

\begin{figure}
\begin{center}
\includegraphics{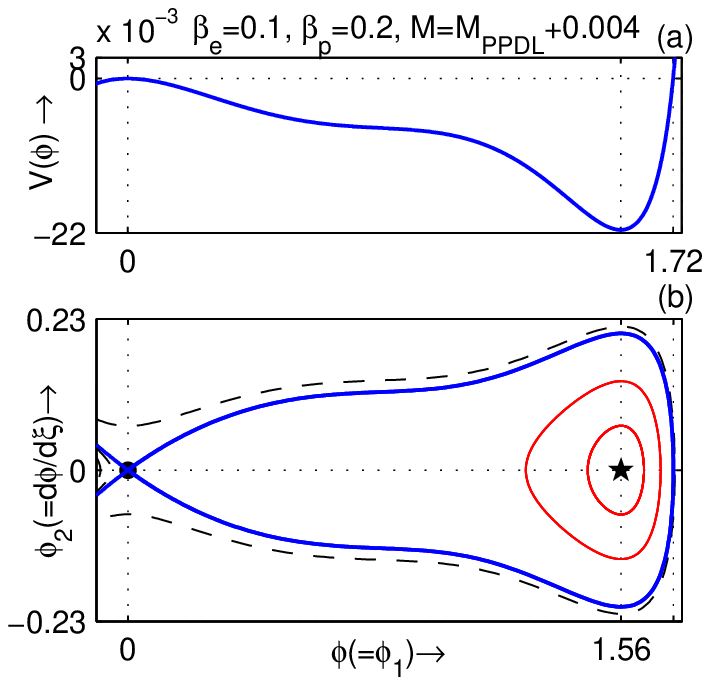}
  \caption{\label{pp1_supersoliton_epsilon=0_pt_004} $V(\phi)$ (on top) and the phase portrait of the system (\ref{phase_portraits}) (on bottom) have been drawn on the same $\phi(=\phi_{1})$-axis for $M=M_{PPDL}+0.004$ when $p=0.07$, $\mu=0.054$, $\beta_{e}=0.1$, $\beta_{p}=0.2$ and $\sigma_{ie}=\sigma_{pe}=0.9$.}
\end{center}
\end{figure}

\begin{figure}
\begin{center}
\includegraphics{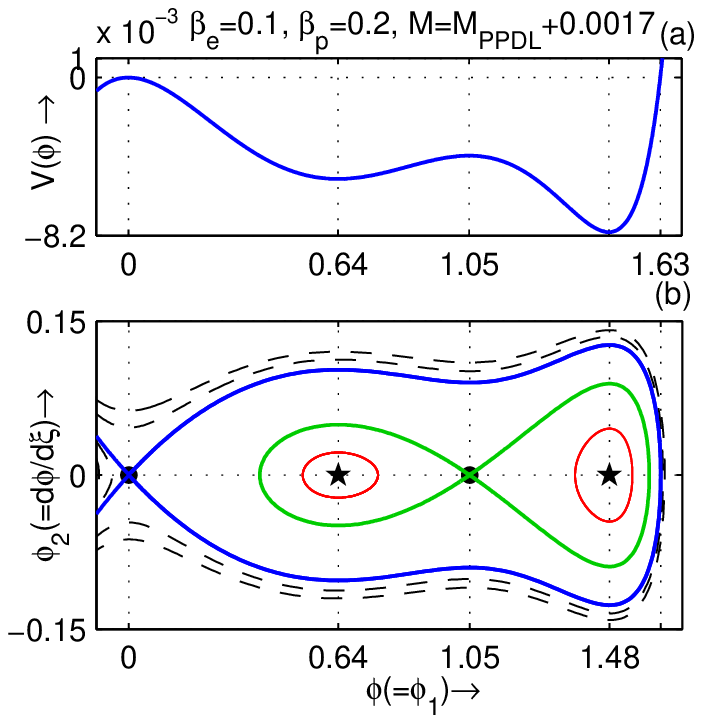}
  \caption{\label{pp1_supersoliton_epsilon=0_pt_0017} $V(\phi)$ (on top) and the phase portrait of the system (\ref{phase_portraits}) (on bottom) have been drawn on the same $\phi(=\phi_{1})$-axis for $M=M_{PPDL}+0.0017$ when $p=0.07$, $\mu=0.054$, $\beta_{e}=0.1$, $\beta_{p}=0.2$ and $\sigma_{ie}=\sigma_{pe}=0.9$.}
\end{center}
\end{figure}

\begin{figure}
\begin{center}
\includegraphics{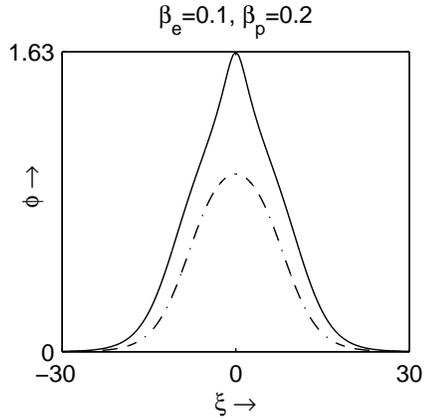}
  \caption{\label{phi_vs_xi_supersoliton} $\phi$ is plotted against $\xi$ for $M=M_{PPDL}+0.0001$ (solid curve) and $M=M_{PPDL}-0.0001$ (dash-dot curve) when $p=0.07$, $\mu=0.054$, $\beta_{e}=0.1$, $\beta_{p}=0.2$ and $\sigma_{ie}=\sigma_{pe}=0.9$.}
\end{center}
\end{figure}

\begin{figure}
\begin{center}
\includegraphics{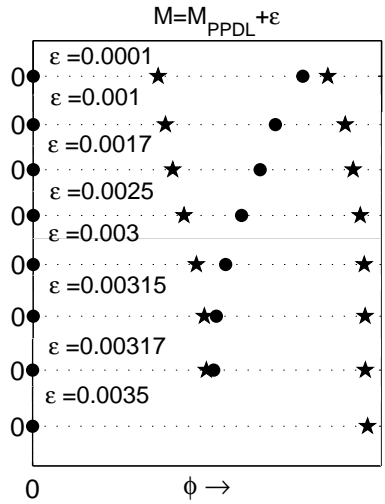}
  \caption{\label{saddle_points} Saddle points (small solid circles) and the equilibrium points other than saddle points (small solid stars) for the system (\ref{phase_portraits}) have been drawn on the $\phi$-axis for different values of the Mach number when $p=0.07$, $\mu=0.054$, $\beta_{e}=0.1$, $\beta_{p}=0.2$ and $\sigma_{ie}=\sigma_{pe}=0.9$.}
\end{center}
\end{figure}

\begin{figure}
\begin{center}
\includegraphics{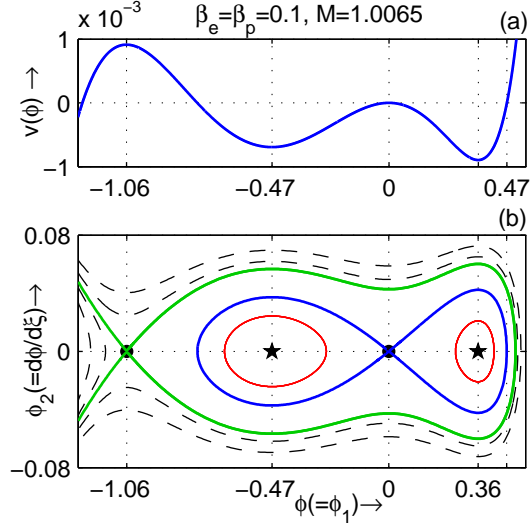}
  \caption{\label{pp1_coexistence} $V(\phi)$ (on top) and the phase portrait of the system (\ref{phase_portraits}) (on bottom) have been drawn on the same $\phi(=\phi_{1})$-axis corresponding to the coexistence of solitons of both polarities for $M=1.0065$ when $p=0.01$, $\mu=0.2$, $\beta_{e}=\beta_{p}=0.1$ and $\sigma_{ie}=\sigma_{pe}=0.9$.}
\end{center}
\end{figure}

\begin{figure}
\begin{center}
\includegraphics{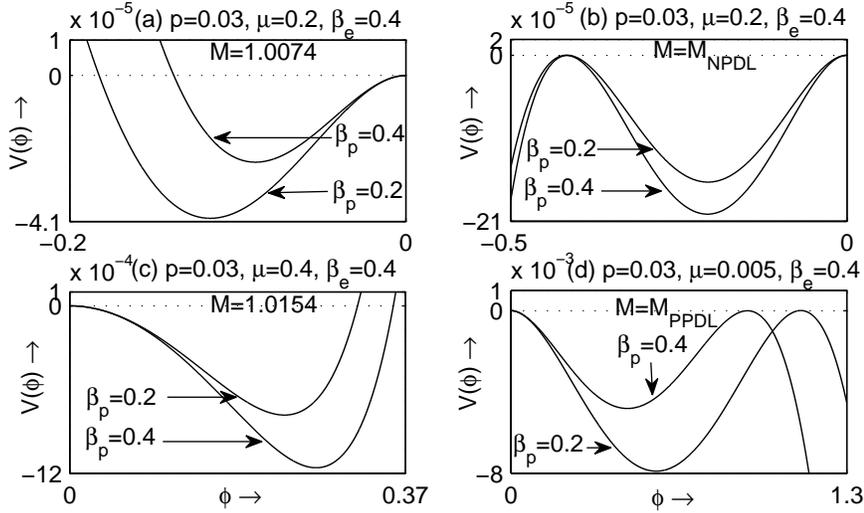}
  \caption{\label{fig_phi_vphi_new_new} In each figure $V(\phi)$ is plotted against $\phi$ for different values of parameters as shown in the figures with $\sigma_{ie}=\sigma_{pe}=0.9$ whenever (a)$M_{c} < M < M_{NPDL}$, (b)$M=M_{NPDL}$, (c)$M_{c} < M \leq M_{max}$ and (d)$M=M_{PPDL}$.}
\end{center}
\end{figure}

A clear scenario of the DIA solitary structures observed in the present investigation has been discussed with the help of phase portraits of the dynamical system corresponding to the DIA solitary structures, giving a special emphasis on the formation of supersolitons. The limitations of the formation of supersolitons have been pointed out through phase portraits. Figure \ref{saddle_points} clearly indicates the limitation for the formation of PPSSs. The same figure (figure \ref{saddle_points}) and the existence of at least one PPSW after the formation of the PPDL (figure \ref{pp1_supersoliton_epsilon=0_pt_004}) indicate the existence of a sequence of supersolitons. Again, there is always a jump type discontinuity between the amplitudes of solitons before and after the formation of double layer (figure \ref{phi_vs_xi_supersoliton}). In our earlier paper \cite{paul2016}, we claimed that the occurrence of soliton after the formation of double layer confirms the existence of supersoliton of same polarity. But we were unable to explain the transition from supersoliton to soliton. In the present paper, with the help of the phase portraits, we have explained the transition process, viz.,  soliton $\to$ double layer $\to$ supersoliton $\to$ soliton for increasing values of Mach number. Irrespective of the model concerned, this transition phenomenon, viz., soliton $\to$ double layer $\to$ supersoliton $\to$ soliton holds good according to the law as described in figure \ref{saddle_points}. The figures, viz., figure \ref{pp1_soliton_epsilon=0_pt_001} - figure \ref{pp1_supersoliton_epsilon=0_pt_0017} and figure \ref{saddle_points} may help to understand the above mentioned transition process. We hope that these discussions develop the theory of supersolitons.

Although  Singh \& Lakhina \cite{singh15} mentioned that there is no direct evidence for the existence of supersolitons in both space and laboratory plasma but in future, next generation satellite expeditions may be able to distinguish the signature for the existence of supersolitons.

\noindent \textbf{Acknowledgements :} One of the authors (Ashesh Paul) is thankful to the Department of Science and Technology, Govt. of India, INSPIRE Fellowship Scheme for financial support.



\newpage

\end{document}